\documentclass[10pt,english,aps,prd,nofootinbib,superscriptaddress,preprintnumbers]{revtex4-2}
\usepackage[T1]{fontenc}
\usepackage[latin9]{inputenc}
\usepackage{textcomp}
\usepackage{amsmath}
\usepackage{amssymb}
\usepackage{graphicx}

\makeatletter

\providecommand{\tabularnewline}{\\}

\def\Mpl{M_{\rm P}}

\usepackage{babel}

\@ifundefined{showcaptionsetup}{}{%
 \PassOptionsToPackage{caption=false}{subfig}}
\usepackage{subfig}
\makeatother

\usepackage{babel}
\begin{document}
\title{Minimal theory of massive gravity and constraints on the graviton
mass}
\author{Antonio De Felice}
\affiliation{Center for Gravitational Physics, Yukawa Institute for Theoretical
Physics, Kyoto University, 606-8502, Kyoto, Japan}
\author{Shinji Mukohyama}
\affiliation{Center for Gravitational Physics, Yukawa Institute for Theoretical
Physics, Kyoto University, 606-8502, Kyoto, Japan}
\affiliation{Kavli Institute for the Physics and Mathematics of the Universe (WPI),
The University of Tokyo, Kashiwa, Chiba 277-8583, Japan}
\author{Masroor C.\ Pookkillath}
\affiliation{Center for Gravitational Physics, Yukawa Institute for Theoretical
Physics, Kyoto University, 606-8502, Kyoto, Japan}
\preprint{YITP-21-102, IPMU21-0060}
\date{\today}
\begin{abstract}
The Minimal theory of Massive Gravity (MTMG) is endowed non-linearly
with only two tensor modes in the gravity sector which acquire a non-zero
mass. On a homogeneous and isotropic background the theory is known
to possess two branches: the self-accelerating branch with a phenomenology
in cosmology which, except for the mass of the tensor modes, exactly
matches the one of $\Lambda$CDM; and the normal branch which instead
shows deviation from General Relativity in terms of both background
and linear perturbations dynamics. For the latter branch we study
using several early and late times data sets the constraints on today's
value of the graviton mass $\mu_{0}$, finding that $(\mu_{0}/H_{0})^{2}=0.119_{-0.098}^{+0.12}$
at 68\% CL, which in turn gives an upper bound at 95\% CL as $\mu_{0}<8.4\times10^{-34}$
eV. This corresponds to the strongest bound on the mass of the graviton
for the normal branch of MTMG. 
\end{abstract}
\maketitle

\section{Introduction}

These years have shown several unexpected results from the experimental
side \cite{2020,Riess_2019,Freedman:2021ahq,KiDS1000,descollaboration2021dark}.
Probably the most important one is the discovery of gravitational
waves \cite{LIGOScientific:2016aoc}. These modes are indeed showing
the tensorial nature of the ripples of the spacetime, as correctly
predicted by Einstein. This new discovery led in turn to a revolution
also into the research of the predictions made by several theoretical
models to the speed of propagation of gravitational waves \cite{LIGOScientific:2017zic}.
In particular, the fact that we have seen two neutron stars merging
with each other, gave us the chance to probe the speed of propagation
for such waves compared to the one of electromagnetic waves together
with the possibility of determining $H_{0}$ \cite{LIGOScientific:2017vwq,LIGOScientific:2019zcs}.
The measurement gave strong constraints on the propagation of gravitational
waves. Although the measurement was regarding small redshifts ($z<0.05$),
nonetheless the gravitational waves had to be traveling through non
trivial backgrounds (the wave had to travel at least the spacetime
around the source, the one of the galaxy to which the source belongs,
the inter-galactic region between the source-galaxy and the Milky
Way, and finally the spacetime of our galaxy and the one of our solar
system, see e.g. \cite{Domenech:2018vqj}. This kinematical observable
pin-pointed the speed of propagation for such waves very close to
the speed of light. In fact, several scalar tensor theories have been
ruled out as predicting a speed of propagation which significantly
differs today from unity \cite{Creminelli:2017sry}.

As a matter of fact, the discovery of the gravitational waves led
necessarily to the search for another fundamental property of theirs:
what is the value of the mass appearing in the dispersion relation for such spacetime ripples? In order
to be consistent with the multi-messenger measurement discussed above,
the mass cannot be too large as this would change the speed of propagation
considerably \cite{theligoscientificcollaboration2020tests,de_Rham_2017}.
General Relativity (and many other theories too) predicts that the
tensor modes are massless. This is indeed consistent with the usual
picture of gravity being a force with an infinite range. However,
without assuming any theoretical prior, we could be addressing, just
using any data at hand, the issue of determining the value of the
mass for gravitational waves.

Although this phenomenological approach is well motivated, still one
would find it more convincing if, after all, there exists a sensible
theory which allows for a non-zero value for the mass of gravitational
waves. If not, even the point of searching would look a bit philosophical,
or at most, it would be an approach which is still incomplete as waiting
for a good theory to be proposed. Fortunately, a theory which allows
for a non-zero mass for the tensorial gravitational waves and at the
same time which is not immediately ruled out by either theoretical
or experimental constraints does exist \cite{DeFelice_2015}. This
theory, the minimal theory of massive gravity (MTMG), is a modification
of the massive gravity theory introduced in \cite{de_Rham_2011} (dRGT).
Compared to dRGT, MTMG introduces a modification by breaking 4D Lorentz-invariance
and removing through appropriate non-linear constraints the (three)
modes which otherwise make the cosmological solutions of dRGT unstable/strongly
coupled \cite{dRGT:ghost}.

MTMG (as well as dRGT) allows for the existence of two different branches,
namely: 1) the self-accelerating branch and 2) the normal branch.
The self-accelerating branch shows exactly the same phenomenology
as $\Lambda$CDM both at the level of the homogeneous and isotropic
background and at the level of linear perturbation theory for scalar
and vector modes \cite{Felice_2016}. The only difference is the presence
of a mass for the tensor modes which does not modify the linear growth
of structures. This branch then shares the same set of observational/experimental constraints
with $\Lambda$CDM.

The normal branch of MTMG is a different beast. It gives a phenomenology
of linear scalar perturbation which is different from $\Lambda$CDM so that
it can give rise to new and interesting constraints from the data
\cite{DeFelice:2016ufg,Bolis_2018}. On top of that, without introducing
any new propagating degrees of freedom, the normal branch of MTMG allows the presence of
a dynamical dark energy component, and such a dynamics can be set
by appropriately choosing the fiducial metric 
to match essentially any desired profile. Recently the simplest
possible dynamics for the normal branch (that reproduces the background
dynamics of $\Lambda$CDM) has been studied in \cite{deAraujo:2021cnd}
and led to the surprising feature that the Planck data tend to set
the squared mass of the tensor modes, $\mu^{2}$, to be a positive
quantity (at 1-sigma). In fact, although this result was consistent
with late time data only, still the data were allowing a large region
of $\mu^{2}<0$ (i.e.\ the graviton could be a ultralight tachyonic
field).

In this work, we seek constraints on the graviton mass in the presence
of a background \emph{dynamical} component that is intrinsically and
generically present in the normal branch of MTMG. We will introduce
five additional new parameters (compared to $\Lambda$CDM) for the
model, but as we will see later on, joining all the chosen data sets
will provide very strong constraints on the (time-dependent) mass
of graviton. In particular we find that in MTMG, the considered data
sets (to be described in more detail in the following, including Planck
2018) do not show internal tensions and all together lead to a bound
on today's value for the mass of the graviton as $\mu_{0}^{2}=2.5{}_{-4.8}^{+4.5}\times10^{-67}\ {\rm eV}^{2}$.
As far as we know, this result is the strongest constraint we have
on the mass of the graviton for the normal branch of MTMG.

\section{The theory}

The theory we want to discuss here is the minimal theory of massive
gravity, MTMG. This theory has been constructed as to have only two
degrees of freedom (the gravitational waves) in the gravity sector
and as to share the same background dynamics as dRGT. In order to
achieve these goals we adopt the usual ADM formalism and consider
the physical lapse function $N$, the physical shift vector $N^{i}$
and the physical three dimensional metric $\gamma_{ij}$ as basic
variables. We also need to choose a fiducial lapse function and a
fiducial three dimensional metric which in the unitary gauge corresponds
to given external fields 
\begin{equation}
M\,,\quad\tilde{\gamma}_{ij}\,,
\end{equation}
together with another three dimensional external field $\tilde{\zeta}^{i}{}_{j}$,
which is related to the rate of change of the vielbein forming $\tilde{\gamma}_{ij}$.
(In the following sections these three external fields will be chosen
to be functions of time only, as to be able to have a homogeneous
description of our universe at large scales.)

We can then introduce the tensor $\mathcal{K}^{i}{}_{j}$ defined
by 
\begin{equation}
\mathcal{K}^{i}{}_{l}\mathcal{K}^{l}{}_{j}=\tilde{\gamma}^{il}\gamma_{lj}\,,
\end{equation}
where $\tilde{\gamma}^{il}$ is the inverse of $\tilde{\gamma}_{ij}$,
that is $\tilde{\gamma}^{il}\tilde{\gamma}_{lj}=\delta^{i}{}_{j}$.
Then we can also introduce the inverse tensor $\mathfrak{K}^{i}{}_{j}$
of $\mathcal{K}^{i}{}_{j}$, defined as 
\begin{equation}
\mathfrak{K}^{i}{}_{l}\mathcal{K}^{l}{}_{j}=\delta^{i}{}_{j}=\mathcal{K}^{i}{}_{l}\,\mathfrak{K}^{l}{}_{j}\,.
\end{equation}
Out of these building blocks we can introduce a symmetric 2-tensor
(in three dimensions) as 
\begin{equation}
\Theta^{ij}=\frac{\sqrt{\tilde{\gamma}}}{\sqrt{\gamma}}\,\{c_{1}\,(\gamma^{il}\mathcal{K}^{j}{}_{l}+\gamma^{jl}\mathcal{K}^{i}{}_{l})+c_{2}\,[\mathcal{K}\,(\gamma^{il}\mathcal{K}^{j}{}_{l}+\gamma^{jl}\mathcal{K}^{i}{}_{l})-2\tilde{\gamma}^{ij}]\}+2c_{3}\,\gamma^{ij}\,,
\end{equation}
where $\tilde{\gamma}$ and $\gamma$ are the determinants of $\tilde{\gamma}_{ij}$
and $\gamma_{ij}$ respectively, whereas $\mathcal{K}\equiv\mathcal{K}^{l}{}_{l}$.
Hereafter, $c_{\mu}$ ($\mu\in\{1,\dots,4\}$) are free numerical
constants. As usual, we can introduce the extrinsic curvature tensor
$K_{ij}$ defined as 
\begin{equation}
K_{ij}=\frac{1}{2N}\,(\dot{\gamma}_{ij}-D_{i}N_{j}-D_{j}N_{i})\,,
\end{equation}
where $D_{i}$ is the three dimensional covariant derivative compatible
with $\gamma_{ij}$, that is $D_{m}\gamma_{ij}=0$. Now, we can define
a three dimensional scalar $\mathcal{C}_{0}$ as 
\begin{equation}
\mathcal{C}_{0}=\frac{1}{2}\,m^{2}\,M\,K_{ij}\,\Theta^{ij}-m^{2}M\left\{ \frac{\sqrt{\tilde{\gamma}}}{\sqrt{\gamma}}\,\{c_{1}\,\tilde{\zeta}+c_{2}\,[(\mathcal{K}\tilde{\zeta}-\mathcal{K}^{i}{}_{l}\tilde{\zeta}^{l}{}_{i})]+c_{3}\,\mathfrak{K}^{i}{}_{l}\tilde{\zeta}^{l}{}_{i}\right\} ,
\end{equation}
where $m$ is a mass-dimension scale (related to the graviton mass),
and we have also named $\tilde{\zeta}\equiv\tilde{\zeta}^{l}{}_{l}$.
It is also necessary to introduce the following three dimensional
2-tensor $\mathcal{C}^{i}{}_{j}$ as 
\begin{equation}
\mathcal{C}^{i}{}_{j}=-m^{2}M\left\{ \frac{\sqrt{\tilde{\gamma}}}{\sqrt{\gamma}}\,\bigl[\tfrac{1}{2}\,(c_{1}+c_{2}\mathcal{K})\,(\mathcal{K}^{i}{}_{j}+\gamma^{il}\mathcal{K}^{m}{}_{l}\gamma_{mj})-c_{2}\,\tilde{\gamma}^{il}\gamma_{lj}\bigr]+c_{3}\,\delta^{i}{}_{j}\right\} .
\end{equation}

We can then introduce the action for MTMG as 
\begin{equation}
S=S_{{\rm pre}}+\frac{\Mpl^{2}}{2}\int d^{4}xN\sqrt{\gamma}\left(\frac{m^{2}}{4}\,\frac{M}{N}\,\lambda\right)^{2}\bigl(\Theta_{ij}\Theta^{ij}-\tfrac{1}{2}\,\Theta^{2}\bigr)-\frac{\Mpl^{2}}{2}\int d^{4}x\sqrt{\gamma}\left[\lambda\mathcal{C}_{0}+(D_{i}\lambda^{j})\mathcal{C}^{i}{}_{j}\right]+S_{{\rm mat}}\,,\label{eqn:fullaction}
\end{equation}
where we have introduced a Lagrange multiplier $\lambda$ together
with another three dimensional vector Lagrange multiplier $\lambda^{i}$.
Furthermore we have defined $\Theta_{ij}\equiv\gamma_{im}\gamma_{jn}\Theta^{mn}$
and $\Theta\equiv\gamma_{ij}\Theta^{ij}$, and $S_{{\rm mat}}$ is
the action for the standard matter fields minimally coupled to the
four-dimensional metric made of the ADM variables ($N$, $N^{i}$,
$\gamma_{ij}$). Finally, $S_{{\rm pre}}$ is given by 
\begin{equation}
S_{{\rm pre}}\equiv S_{{\rm GR}}+\frac{\Mpl^{2}}{2}\,\sum_{n=1}^{4}\int d^{4}x\,\mathcal{S}_{n}\,,
\end{equation}
where 
\begin{eqnarray}
S_{{\rm GR}} & = & \frac{\Mpl^{2}}{2}\int d^{4}x\,N\sqrt{\gamma}\,[{}^{(3)}R+K^{ij}K_{ij}-K^{2}]\,,\\
\mathcal{S}_{1} & = & -m^{2}c_{1}\sqrt{\tilde{\gamma}}\,(N+M\mathcal{K})\,,\\
\mathcal{S}_{2} & = & -\frac{1}{2}\,m^{2}\,c_{2}\sqrt{\tilde{\gamma}}\,(2N\mathcal{K}+M\mathcal{K}^{2}-M\tilde{\gamma}^{ij}\gamma_{ji})\,,\\
\mathcal{S}_{3} & = & -m^{2}\,c_{3}\sqrt{\gamma}\,(M+N\mathfrak{K})\,,\\
\mathcal{S}_{4} & = & -m^{2}\,c_{4}\sqrt{\gamma}\,N\,,
\end{eqnarray}
$^{(3)}R$ is the Ricci scalar for the three dimensional metric $\gamma_{ij}$,
$K^{ij}\equiv\gamma^{im}\gamma^{jn}K_{mn}$, $K\equiv\gamma^{ij}K_{ij}$,
and $\mathfrak{K}\equiv\mathfrak{K}^{l}{}_{l}$.

\section{The cosmological background}

The action (\ref{eqn:fullaction}) is the full action of the theory
which can be studied on any desired background, as long as it is compatible
with the equations of motion for the theory. In the following we will
focus on a homogeneous and isotropic background, as to study the cosmology
for this theory. First of all, let us fix the fiducial metric $\tilde{\gamma}_{ij}=\tilde{a}(t)^{2}\,\delta_{ij}$
(so that $\tilde{\gamma}^{ij}=\tilde{a}^{-2}\,\delta^{ij}$), and,
at the same time, also the fiducial lapse $M=M(t)$, and the rate
of change for the fiducial vielbein $\tilde{\zeta}^{i}{}_{j}=\frac{\dot{\tilde{a}}}{M\,\tilde{a}}\,\delta^{i}{}_{j}$.
Notice that having adopted the unitary gauge these fiducial variables
come as purely background quantities. From the viewpoint of the physical
sector, these correspond to external fields which explicitly break
Lorentz invariance at the cosmological scale.

As for the remaining variables we will first set up the physical lapse,
shift and 3D metric as follows: 
\begin{eqnarray}
N & = & N(t)\,(1+\alpha)\,,\\
N_{i} & = & N(t)\,\partial_{i}\chi\,,\\
\gamma_{ij} & = & a(t)^{2}\,[\delta_{ij}(1+2\zeta)+2a^{-2}\partial_{i}\partial_{j}E]\,.
\end{eqnarray}
Here, we have set not only the background but also the linear perturbation
variables, $\alpha,\chi,\zeta$, and $E$. Since we have imposed the
unitary gauge to hold, we cannot impose any further gauge conditions
on the perturbation variables. On the other hand, we still have a
freedom to re-select an arbitrary monotonic function of the time coordinate
for the background of the temporal Stückelberg field. As a result,
we can freely set, if needed, $N(t)$ to a conveniently chosen positive
function of time, e.g.\ $N(t)=1$ or $N(t)=a(t)$. (If we a priori
fix $N(t)$ in this way then the background equations of motion determines
$M(t)$ instead of $N(t)$. See (\ref{eqn:M-normalbranch}) below.)
Now we can find the solution to the equation $\mathcal{K}^{i}{}_{l}\mathcal{K}^{l}{}_{j}=\tilde{\gamma}^{il}\gamma_{lj}$,
order by order in perturbations, giving the fact that at the lowest
order (i.e.\ on the background) we have $\mathcal{K}^{i}{}_{j}=(a/\tilde{a})\,\delta^{i}{}_{j}=\delta^{i}{}_{j}/X$.
Here, we have introduced the background variable $X=X(t)$ defined
as $\tilde{a}=X\,a$. We are now able to write down $\mathcal{S}_{1}$,
$\mathcal{S}_{2}$ and $\mathcal{S}_{4}$. Along the same line, on
finding the inverse of $\mathcal{K}^{i}{}_{j}$, namely $\mathfrak{K}^{i}{}_{j}$,
we can write down also $\mathcal{S}_{3}$. After introducing 
\begin{eqnarray}
\lambda & = & \lambda(t)+\delta\lambda\,,\\
\lambda^{i} & = & \frac{1}{a^{2}}\,\delta^{ij}\partial_{j}\delta\lambda_{f}\,,
\end{eqnarray}
we have all the remaining building blocks which form the full action
of the theory. Here, we have focused our attention only to the background
and scalar perturbation modes. As for the remaining ones, vector modes
dynamics is the same as GR (see e.g.\ \cite{Felice_2016}), whereas
the tensor modes will get a massive dispersion relation, which will
be described, for our convenience, later on.

We can introduce the matter fields in the usual ADM formalism without
fixing any gauge. Just to give an example we will write here a perfect
fluid component (labeled with a variable $I$), whose Schutz-Sorkin
action~\cite{Schutz:1977df,Pookkillath:2019nkn} reads 
\begin{equation}
S_{{\rm pf}}^{(I)}=-\int d^{4}x\sqrt{-g}[\rho_{I}(n_{I})+J_{I}^{\mu}\partial_{\mu}\varphi_{I}]\,.
\end{equation}
Here, $n_{I}\equiv\sqrt{-J_{I}^{\mu}J_{I}^{\nu}g_{\mu\nu}}$, and
$J_{I}^{\mu}$ form the components of a time-like four-vector, out
of which we have the normalized 4-velocity of the fluid $u_{I}^{\mu}\equiv J_{I}^{\mu}/n$.
Then we write 
\begin{eqnarray}
J_{I}^{0} & = & \frac{J_{I}^{0}(t)}{N(t)}\,(1+\delta j_{0}^{I})\,,\\
J_{I}^{i} & = & \frac{1}{a^{2}}\,\delta^{ij}\partial_{j}\delta j_{I}\,,\\
\varphi_{I} & = & \varphi_{I}(t)+\delta\varphi_{I}\,.
\end{eqnarray}
The background equations of motion for the fluid impose that $J_{I}^{0}(t)=\mathcal{N}_{I}/a^{3}$,
where $\mathcal{N}_{I}={\rm constant}$, together with $\varphi_{I}=-\int^{t}N(t')\,\frac{\partial\rho_{I}}{\partial n_{I}}\,dt'$.
Having fixed the matter fields, whose background equations of motion
do not get any modification (after all we are only changing the gravity
sector), we are ready to move on to the remaining modified Einstein
equations and the additional constraints introduced in MTMG.

One property of MTMG is that on the cosmological background the equations
of motion lead, as a unique solution, to $\lambda(t)=0$. In this
case the modified Friedmann equation reads 
\begin{eqnarray}
3\Mpl^{2}H^{2} & = & \rho_{X}+\sum_{I}\rho_{I}\,,\\
\rho_{X} & \equiv & \frac{1}{2}\,m^{2}\Mpl^{2}\,(c_{1}X^{3}+3c_{2}X^{2}+3c_{1}X+c_{4})\,.\label{eqn:modifiedFriedmanneq}
\end{eqnarray}
The constraint introduced in MTMG leads to the following equation
of motion 
\begin{equation}
\mathcal{E}_{\lambda}=(c_{1}X^{2}+2c_{2}X+c_{3})\left(\frac{\dot{X}}{N}+HX-H\,\frac{M}{N}\right)=0\,.\label{eqn:FLRW-constraint}
\end{equation}
As in dRGT, the factorized structure of this equation allows for the
presence of two distinct branches of solutions~\cite{Gumrukcuoglu:2011ew}.
In particular the so-called self-accelerating branch, defined by setting
the first factor to vanish, leads to a quadratic algebraic equation
for $X$, which implies $X={\rm constant}$. As a consequence, on
this branch $\rho_{X}={\rm constant}$, and the dynamics of the cosmological
background is the same as the one for $\Lambda$CDM.

On the other hand, for the so-called normal branch we solve $\mathcal{E}_{\lambda}=0$
by setting the second factor in (\ref{eqn:FLRW-constraint}) to vanish.
This in turn, assuming $H\neq0$, leads to 
\begin{equation}
M=\frac{\dot{X}}{H}+N\,X\,.\label{eqn:M-normalbranch}
\end{equation}
Since now on, we will consider this branch since the self-accelerating
branch is indistinguishable from $\Lambda$CDM (even at the level
of perturbations, as far as scalar and vector perturbations are concerned~\footnote{The tensor modes acquire a non-zero mass and thus behave differently
from $\Lambda$CDM. However, if we consider the graviton mass term
as the origin of the accelerated expansion of the present universe,
then the mass of the tensor modes will be of order of $H_{0}$, well
below the sensitivity of experiments. On the other hand, if we do
not make this assumption then the strongest bound on the graviton
mass in the self-accelerating branch is the bound from observations
of gravitational waves, which is much weaker. }). Notice that we have fixed $M(t)$ as in (\ref{eqn:M-normalbranch}),
but still $X(t)$ is free, because $\tilde{a}(t)$, the scale factor
of the fiducial metric, has to be understood as a given function of
time (or as a fixed function of the temporal Stückelberg field in
the covariant formulation), that can be freely specified as a part
of the definition of the theory.

We can rewrite then the modified second Einstein equation as 
\begin{eqnarray}
2\Mpl^{2}\,\frac{\dot{H}}{N} & = & -\sum_{I}(\rho_{I}+P_{I})-(\rho_{X}+P_{X})\,,\label{eqn:dotH}\\
\rho_{X}+P_{X} & = & -\frac{1}{2}\,m^{2}\Mpl^{2}\,\frac{\dot{X}}{NX}\,(c_{1}X^{2}+2c_{2}X+c_{3})\,.\label{eqn:rhoX+PX}
\end{eqnarray}
As a direct consequence of these equations, in the normal branch,
in general we should expect a time-dependent, i.e.\ dynamical component
$\rho_{X}$ whose dynamics can be given a priori. In other words,
although in the gravity sector there are no propagating degrees of
freedom beside the gravitational tensor modes, still the background
has non-trivial dynamics, in general different from the one of $\Lambda$CDM\footnote{In the normal branch, as a particular case, one can choose $X(t)$
to be a constant, that is $\dot{X}=0$, and then the background becomes
indistinguishable from the one of $\Lambda$CDM. Even in this case,
perturbations will still behave differently from the standard model
of cosmology. }. What we intend to do in the present paper is to find constraints
on the mass of the graviton in the normal branch allowing non-trivial
background dynamics for $X(t)$, or, equivalently, for $\rho_{X}$.

\subsection{Background dynamics}

Given the functional freedom of choosing $\rho_{X}$, we will consider
the normal branch of MTMG with a background which deviates only slightly
from the one of $\Lambda$CDM. In particular, let us consider a background
which interpolates two $\Lambda$CDM's with slightly different cosmological
constants. This interpolation is defined by giving an explicit form
for the Hubble expansion rate in terms of the redshift $z$ as follows
\begin{equation}
\frac{H^{2}}{H_{0}^{2}}=\frac{H_{\Lambda{\rm {CDM}}}^{2}(z)}{H_{0}^{2}}+\left[f(z)-1\right]\Delta+\mathcal{O}(\Delta^{2})\,,\label{eqn:H-ansatz}
\end{equation}
where 
\begin{eqnarray}
\frac{H_{\Lambda{\rm {CDM}}}^{2}(z)}{H_{0}^{2}} & = & \Omega_{m0}(1+z)^{3}+\Omega_{r0}(1+z)^{4}+1-\Omega_{m0}-\Omega_{r0}\,,\\
f(z) & = & \frac{1+\tanh\frac{A_{2}-z}{A_{2}A_{3}}}{1+\tanh A_{3}^{-1}}\,.
\end{eqnarray}
This form was motivated by a recent work in the context of minimally
modified gravity~\cite{DeFelice:2020cpt}. The exact $\Lambda$CDM
limit takes place when $\Delta\to0$. Also notice that at low redshifts
$0\leq z\ll A_{2}$, we have $\left[f(z)-1\right]\Delta\to0$ and
thus $H\simeq H_{\Lambda{\rm {CDM}}}(z)$. On the other hand, when
$z\gg A_{2}$ (high redshifts), then 
\begin{equation}
\frac{H^{2}}{H_{0}^{2}}=\frac{H_{\Lambda{\rm {CDM}}}^{2}(z)}{H_{0}^{2}}-\Delta+\mathcal{O}(\Delta^{2})\,,
\end{equation}
which corresponds to a shift of the cosmological constant (the sign
of $\Delta$ will be fixed by the data).

The modified Friedmann equation of MTMG, in the presence of some radiation
and dust components is written as 
\begin{equation}
3\Mpl^{2}H^{2}=\rho_{m}+\rho_{r}+\rho_{X}\,,\label{eq:mtmg-fr1}
\end{equation}
where $\rho_{X}$ is given by (\ref{eqn:modifiedFriedmanneq}). Now,
in order to realize the behavior of $H(z)$ shown in (\ref{eqn:H-ansatz}),
let us consider $X$ of the form 
\begin{equation}
X=X(z)=1+(A_{1}-1)\,f(z)\,,\quad f(z)=\frac{1+\tanh\frac{A_{2}-z}{A_{2}A_{3}}}{1+\tanh(A_{3}^{-1})}\,.\label{eqn:X(z)}
\end{equation}
(The value of $\Delta$ will be given in terms of $A_{1}$ as shown
in (\ref{eqn:Delta-A1}) below.) In this case $X$ will show a transition
between two different constant values, namely from $X=1$ for large
redshifts to $X=A_{1}$. Indeed, if we require that $A_{2}>0$ (for
the MC sampling we assume $A_{2}>0.12$ for the reason that will be
explained later) and that $0<A_{3}\ll1$, then the transition happens
in the past. As a result, $\rho_{X}$ will also interpolate two constants.
Notice that this transition is smooth for both $X$ and $\rho_{X}$,
taking place at around $z\approx A_{2}$. Then we have at all times
\begin{equation}
\frac{H^{2}}{H_{0}^{2}}=\Omega_{m0}(1+z)^{3}+\Omega_{r0}(1+z)^{4}+\frac{m^{2}}{6H_{0}^{2}}\,(c_{1}X^{3}+3c_{2}X^{2}+3c_{3}X+c_{4})\,,
\end{equation}
as a given function of $z$. We can redefine the parameters of the
theory as 
\begin{equation}
\bar{c}_{\mu}=c_{\mu}\,\frac{m^{2}}{H_{0}^{2}}\,,\qquad\mu\in\{1,\dots,4\}\,,
\end{equation}
so that 
\begin{equation}
\frac{H^{2}}{H_{0}^{2}}=\Omega_{m0}(1+z)^{3}+\Omega_{r0}(1+z)^{4}+\frac{1}{6}\,(\bar{c}_{1}X^{3}+3\bar{c}_{2}X^{2}+3\bar{c}_{3}X+\bar{c}_{4})\,,\label{eqn:H-X}
\end{equation}
where $X$ is given by (\ref{eqn:X(z)}). In the following, we will
make use of the following variables 
\begin{eqnarray}
\varrho_{I} & \equiv & \frac{\rho_{I}}{3\Mpl^{2}}\,,\quad p_{I}\equiv\frac{P_{I}}{3\Mpl^{2}}\,,\\
\varrho_{X} & \equiv & \frac{\rho_{X}}{3\Mpl^{2}}=(H_{0}^{2}/6)\,(\bar{c}_{1}X^{3}+3\bar{c}_{2}X^{2}+3\bar{c}_{3}X+\bar{c}_{4})\,,\\
p_{X} & \equiv & \frac{P_{X}}{3\Mpl^{2}}=-\frac{1}{3}\frac{H_{0}^{2}}{2}\,X(\bar{c}_{1}X^{2}+2\bar{c}_{2}X+\bar{c}_{3})\,\frac{\dot{X}}{NHX}-\varrho_{X}\,,\nonumber \\
 & = & -\frac{1}{3}\,H_{0}^{2}\,\theta\,\epsilon_{X}-\varrho_{X}\,,\\
\theta(t) & \equiv & \frac{1}{2}\,X(\bar{c}_{1}X^{2}+2\bar{c}_{2}X+\bar{c}_{3})\,, \label{theta_bg}\\
\epsilon_{X} & \equiv & \frac{\dot{X}}{NHX}=-(1+z)\frac{1}{X}\,X_{,z}\,,
\end{eqnarray}
where $X_{,z}=dX/dz$. From the second Einstein equation we also find
(\ref{eqn:dotH})-(\ref{eqn:rhoX+PX}), which can be rewritten as
\begin{equation}
\frac{\dot{H}}{N}=-\frac{3}{2}(\varrho_{X}+p_{X})-\frac{3}{2}\,\sum_{I}(\varrho_{I}+p_{I})\,.
\end{equation}

On defining 
\begin{equation}
\Omega_{{\rm DE},0}\equiv1-\Omega_{m0}-\Omega_{r0}\,,
\end{equation}
we find that for $z=0$ 
\begin{equation}
\Omega_{{\rm DE},0}=\frac{1}{6}\,(\bar{c}_{1}A_{1}^{3}+3\bar{c}_{2}A_{1}^{2}+3\bar{c}_{3}A_{1}+\bar{c}_{4})\,,
\end{equation}
which can be used to rewrite $\bar{c}_{4}$ as 
\begin{equation}
\bar{c}_{4}=6\Omega_{{\rm DE},0}-(\bar{c}_{1}A_{1}^{3}+3\bar{c}_{2}A_{1}^{2}+3\bar{c}_{3}A_{1})\,.
\end{equation}
On the other hand at high redshifts $X\approx1$, we have $\rho_{X}\to{\rm constant}$
and thus matching (\ref{eqn:H-X}) (with $X$ given by (\ref{eqn:X(z)}))
to (\ref{eqn:H-ansatz}) results in 
\begin{equation}
\frac{1}{6}\,(\bar{c}_{1}+3\bar{c}_{2}+3\bar{c}_{3})+\frac{1}{6}\bar{c}_{4}=\Omega_{{\rm DE},0}-\Delta+\mathcal{O}(\Delta^{2})\,,
\end{equation}
which can be used in order to write down $\Delta$ in terms of $A_{1}$
and $\bar{c}_{i}$ ($i\in\{1,2,3\}$) as 
\begin{equation}
\Delta=\frac{1}{6}[\bar{c}_{1}(A_{1}^{3}-1)+3\bar{c}_{2}(A_{1}^{2}-1)+3\bar{c}_{3}(A_{1}-1)]\,.\label{eqn:Delta-A1}
\end{equation}
This vanishes either when $A_{1}=1$ or when $\bar{c}_{i}\to0$. In
other words, the exact $\Lambda$CDM background is recovered by setting
either $A_{1}=1$ or $\bar{c}_{i}=0$ ($i\in\{1,2,3\}$). We also
have 
\begin{equation}
\frac{M}{N}=\frac{\dot{X}}{HN}+X=-(1+z)\,X_{,z}+X\,,
\end{equation}
and the squared mass of the graviton, as we shall see later on, reads
\begin{equation}
\mu^{2}=\frac{1}{2}\,m^{2}\,X\left[c_{2}X+c_{3}+\frac{M}{N}\,(c_{1}X+c_{2})\right].
\end{equation}
If we assume that $A_{2}>0$ and $0<A_{3}\ll1$, then around today
we find, since $X_{,z}\approx0$, that $M/N=X=A_{1}$ so that 
\begin{equation}
\theta_{0}\equiv\frac{\mu_{0}^{2}}{H_{0}^{2}}=\frac{1}{2}\,A_{1}\,[\bar{c}_{1}A_{1}^{2}+2\bar{c}_{2}A_{1}+\bar{c}_{3}]\,,
\end{equation}
which can be used to find today's value of the mass of the graviton
in terms of $A_{1}$ and $c_{i}$. In the $\Lambda$CDM limit this
mass vanishes.

\subsection{Independent model parameters}

Then we choose $A_{1},A_{2},\Omega_{{\rm DE},0},\bar{c}_{i}$ ($i\in\{1,2,3\}$)
as the independent free parameters to run MC sampling. In principle
we could also count the parameter $A_{3}$ (which sets the speed of
transition). However, we have checked that the bestfit to the data
does not depend on $A_{3}$, so that we can safely fix it to a small
value as $A_{3}=10^{-3}$. On top of these $5$ ($+1$) parameters
we also have the standard matter parameters. Then, as already stated
above, one can find the following derived parameters 
\begin{eqnarray}
\theta_{0} & = & \frac{1}{2}A_{1}^{3}\bar{c}_{1}+A_{1}^{2}\bar{c}_{2}+\frac{1}{2}A_{1}\bar{c}_{3}\,,\\
\Delta & = & \left(A_{1}^{3}-1\right)\frac{\bar{c}_{1}}{6}+\left(A_{1}^{2}-1\right)\frac{\bar{c}_{2}}{2}+\left(A_{1}-1\right)\frac{\bar{c}_{3}}{2}\,,\\
\bar{c}_{4} & = & -A_{1}^{3}\bar{c}_{1}-3A_{1}^{2}\bar{c}_{2}-3A_{1}\bar{c}_{3}+6\Omega_{{\rm DE},0}\,,
\end{eqnarray}
where 
\begin{equation}
\Omega_{{\rm DE},0}\equiv1-\Omega_{m0}-\Omega_{r0}\,.
\end{equation}

It should be noted that $A_{1}>0$, because $X$ is required to be
positive. Otherwise the fiducial metric would have a vanishing scale
factor. Furthermore, in order to prevent a rapid change from taking
place only around today, we assume $A_{2}$ to be larger than 0.12.
Flat priors with a sufficiently wide range are instead given to the
$\bar{c}_{i}$ parameters.

\section{Cosmological Perturbations}

Let us first define all the perturbation variables which enter in
the theory, keeping in mind we have already fixed a gauge for the
perturbations, namely the unitary gauge. This study has been already
performed in \cite{deAraujo:2021cnd}, and thus we will only summarize
here the results. In any case, the equations of motion to be inserted
in the Boltzmann solver appear here, as far as we know, for the first
time in the literature. Since the vector perturbations dynamics have
been shown to be exactly equal to the $\Lambda$CDM case, we will
not mention them in the following. Instead, as usual, the scalar sector
needs to be explained in more detail.

We have already introduced the variables up to first order in perturbation
theory in the scalar sector. In the following we will find it useful
to consider the following definitions 
\begin{eqnarray}
\eta_{X}(t) & \equiv & \bar{c}_{2}X(t)^{2}+\bar{c}_{3}X(t)+\frac{1}{2}\theta(t)^{2}Y(t)-2\theta(t)\,,\qquad Y(t)\equiv\frac{H_{0}^{2}}{H^{2}}\,,\\
\Gamma & = & \sum_{I}\Gamma_{I}\,,\qquad\Gamma_{I}=\varrho_{I}+p_{I}\,.
\end{eqnarray}
We can now expand up to second order the total action, including the
total matter Lagrangian. One can introduce a phenomenological description
of matter by starting with Lagrangians for perfect fluids and then
adding to them the shear perturbation terms as done in \cite{DeFelice:2020cpt}.
For the Schutz-Sorkin action for perfect fluids, already introduced
above, we have $u_{Ii}=J_{Ii}/n_{I}=g_{i\alpha}J_{I}^{\alpha}/n_{I}$.
This leads to 
\begin{equation}
\delta j_{I}=\frac{\mathcal{N}_{I}}{a^{3}}\,(v_{I}-\chi)\,,
\end{equation}
where we have defined $u_{Ii}=\partial_{i}v_{I}$. Through this field
redefinition, we have now introduced a field which has a simple physical
meaning. The equation of motion for $\delta j_{I}$ can now be used
in order to set the following constraint 
\begin{equation}
\delta\varphi_{I}=\rho_{I,n}\,v_{I}\,,
\end{equation}
which eliminates fields in the matter sector. By expanding the quantity
$\rho_{I}/\rho_{I}(t)-1$ up to first order of perturbations, and
calling such a variable $\delta\rho_{I}/\rho_{I}$, then we have that
\begin{equation}
\delta j_{0}^{I}=\frac{\rho_{I}}{n_{I}\rho_{I,n}}\,\frac{\delta\rho_{I}}{\rho_{I}}-\alpha\,.
\end{equation}
This equation can be used as a field redefinition introducing the
fields $\delta\rho_{I}/\rho_{I}$ in the matter sector, which have
a clear physical meaning.

At this level we can find the equations of motion for all the perturbations
we have in the gravity and matter sectors. For instance, by ``$\mathcal{E}_{\alpha}$''
we will name the equation of motion obtained by taking variation of
the second order action with respect to the field $\alpha$. In this
case we have the following equations of motion: $\mathcal{E}_{\alpha}$,
$\mathcal{E}_{\chi}$, $\mathcal{E}_{\zeta}$, $\mathcal{E}_{E}$,
$\mathcal{E}_{\delta\lambda}$, $\mathcal{E}_{\delta\lambda_{f}}$,
$\mathcal{E}_{\delta\rho_{I}/\rho_{I}}$, $\mathcal{E}_{v_{I}}$.
Each of these expression is required to vanish, being the equations
of motion for the perturbations.

We can now set $N(t)=a(t)$ on the background, but we cannot choose
any gauge fixing for the perturbation fields since we have already
adopted the unitary gauge. Nonetheless, since Lorentz violation is
introduced only in the gravity sector at cosmological scales and the
percolation of the Lorentz violation to the matter sector due to graviton
loops is suppressed by negative powers of $\Mpl^{2}$, any quantities
that can be observed by any probes made of matter fields should be
(either exactly or approximately) gauge-invariant in the sense of
the four-dimensional diffeomorphism. Therefore it is useful to define
gauge invariant perturbation variables as follows 
\begin{eqnarray}
\alpha & = & \psi-\frac{1}{a}\,\dot{\chi}+\frac{1}{a}\,\partial_{t}\!\left[a\partial_{t}\!\left(\frac{E}{a^{2}}\right)\right],\\
\zeta & = & -\phi-H\,\chi+aH\,\partial_{t}\!\left(\frac{E}{a^{2}}\right),\\
\frac{\delta\rho_{I}}{\rho_{I}} & = & \delta_{I}-\frac{\dot{\rho}_{I}}{a\rho_{I}}\,\chi+\frac{\dot{\rho}_{I}}{\rho_{I}}\,\partial_{t}\!\left(\frac{E}{a^{2}}\right),\\
v_{I} & = & -\frac{a}{k^{2}}\,\theta_{I}+\chi-a\,\partial_{t}\!\left(\frac{E}{a^{2}}\right).
\end{eqnarray}
These gauge invariant variables, namely $\psi$, $\phi$, $\delta_{I}$,
$\theta_{I}$ are nothing but the gauge invariant variables which
define the longitudinal-gauge variables, including the Bardeen
potentials~\cite{Bardeen:1980kt}. Here $\theta$ with index $I$
represents the field associated with matter velocity and is a
perturbation variable. The index $I$ refers for fluid components like,
radiation $r$, cold dark matter $c$, etc. It is different from
$\theta$ that appeared in the background variable associated with the
pressure from the minimal massive gravity (\ref{theta_bg}).

As for the matter equations of motion, they exactly reduce to the
$\Lambda$CDM case, namely 
\begin{eqnarray}
\dot{\delta}_{I} & = & -3aH(c_{sI}^{2}-w_{I})\,\delta_{\mathrm{I}}-(1+w_{I})\,\theta_{I}+3(1+w_{I})\,\dot{\phi}\,,\\
\dot{\theta}_{I} & = & aH(3c_{sI}^{2}-1)\,\theta_{I}+k^{2}\psi+\frac{c_{sI}^{2}k^{2}}{1+w_{I}}\,\delta_{I}-k^{2}\sigma_{I}\,,
\end{eqnarray}
where $c_{sI}^{2}=(\partial p_{I}/\partial\rho_{I})_{s}$, and $w_{I}=p_{I}/\rho_{I}$.
This standard result in the matter sector is actually expected as
MTMG does not modify matter Lagrangians.

However, we should expect deviations when we consider the perturbed
Einstein equations as we will show in the following. In fact, the
equation of motion $E_{\delta\lambda_{V}}=0$, sets 
\begin{equation}
\chi=-\frac{1}{H}\,\phi+a\,\partial_{t}\!\left(\frac{E}{a^{2}}\right)\,,
\end{equation}
or, equivalently $\zeta=0$. We have used one equation and set one
variable. We can consider a linear combination of the form $\mathcal{E}_{\alpha}+3Ha^{2}\,\mathcal{E}_{\chi}/k^{2}$
as to set the field $E$, and we can set $\delta\lambda$ by using
$\mathcal{E}_{\chi}$. On using now $\mathcal{E}_{E}$, we find 
\begin{equation}
\mathcal{E}_{1}\equiv\dot{\phi}+\frac{3a\theta Y\left(\Gamma-\frac{\epsilon_{X}\left(Y\theta-2\right)H^{2}}{3}\right)}{2H\left(Y\theta-2\right)}\,\phi+Ha\,\psi+\frac{3a^{2}}{k^{2}(Y\theta-2)}\,\sum_{I}\Gamma_{I}\theta_{I}=0\,,
\end{equation}
which is one of the two dynamical equations that we use in the Boltzmann
solver. Another linear combination of the remaining equations of motion,
namely $\mathcal{E}_{\zeta}+3a^{2}\mathcal{E}_{E}/k^{2}$, can be
used to set $\delta\lambda_{f}$. The equation of motion which has
not yet been used is $\mathcal{E}_{\delta\lambda}$, which turns out
to give a relation among $\phi,\theta_{I},\delta_{I}$ as 
\begin{equation}
\mathcal{E}_{0}\equiv\left[\frac{(Y\theta-2)k^{2}}{a^{2}}+\frac{9Y\theta\Gamma}{2}\right]\phi+\frac{9aH(Y\theta-2)}{2k^{2}}\,\sum_{I}\Gamma_{I}\theta_{I}-3\sum_{I}\varrho_{I}\delta_{I}=0\,.\label{eq:eom_0}
\end{equation}
On taking a time derivative of $\mathcal{E}_{0}$, and using $\mathcal{E}_{1}$
in order to remove the $\dot{\phi}$ term, and the matter equations
of motion to remove $\dot{\theta}_{I}$ and $\dot{\delta}_{I}$, we
arrive at another equation of motion, the shear equation of motion,
which can be written as follows. 
\begin{eqnarray}
\mathcal{E}_{2} & \equiv & \psi+\frac{9\,a^{2}}{2k^{2}}\,\sum_{I}\Gamma_{I}\sigma_{I}-\left[1+\frac{3Y\theta\Gamma}{(2Y\theta-4)H^{2}}+\frac{27a^{2}Y\theta\left(\sum_{I}c_{s,I}^{2}\Gamma_{I}-\frac{\Gamma^{2}}{2H^{2}}\right)}{2k^{2}(2-Y\theta)}+\left(1+\frac{9\Gamma\,a^{2}}{2k^{2}}\right)\frac{Y\eta_{X}\epsilon_{X}}{2-Y\theta}\right]\phi\nonumber \\
 &  & +\frac{9a^{3}Y\left(\eta_{X}H^{2}\epsilon_{X}-\frac{3\theta\Gamma}{2}\right)}{2H\,k^{4}\left(Y\theta-2\right)}\sum_{I}\Gamma_{I}\theta_{I}-\frac{9a^{2}Y\theta}{2k^{2}\left(Y\theta-2\right)}\,\sum_{I}c_{s,I}^{2}\,\varrho_{I}\,\delta_{I}=0\,.
\end{eqnarray}
The equations $\mathcal{E}_{1}$ and $\mathcal{E}_{2}$ are the equations
of motion which need to be implemented in the Boltzmann solver. At
this level all the equations of motion have been used and all the
variables have been set, and the independent equations of motion form
a closed system of ODE's in Fourier space. Since MTMG does not add
any new degree of freedom in the gravity (or matter) sector, we are
now ready to study the dynamics of linear perturbations for the normal
branch of MTMG having a fully dynamical non-$\Lambda$CDM background.
Before discussing the numerical results which will set constraints
on such dynamics, we would like to investigate three other issues: 1)
Having these equations of motion, what is the effective gravitational
constant for MTMG? 2) Does this dynamical MTMG background lead to
ghosts? 3) What is the dynamics for the tensor modes?

\subsection{Effective gravitational constant}

In order to give a value for the effective gravitational constant
in the normal branch of MTMG, let us consider as for matter only a
single dust fluid (the introduction of an extra cold baryon fluid
is trivial and will not affect the results). This approximation only
holds evidently at late times when the radiation components can be
neglected. Then the matter equations of motion read 
\begin{eqnarray}
\dot{\delta}_{c} & = & -\theta_{c}+3\,\dot{\phi}\,,\label{eq:dot_delta_c}\\
\dot{\theta}_{c} & = & -aH\,\theta_{c}+k^{2}\psi\,,\label{eq:dot_theta_c}
\end{eqnarray}
as dust has no shear. We can solve now $\mathcal{E}_{2}$ for $\psi$
in terms of $\phi$, and $\theta_{c}$. Now let us solve algebraically
$\mathcal{E}_{1}$ for $\dot{\phi}$ in terms of $\phi$ and $\theta_{c}$
only, having already substituted $\psi$ in it. Then on substituting
this expression for $\dot{\phi}$ in Eq.\ (\ref{eq:dot_delta_c}),
we will have a relation among $\theta_{c}$, $\phi$, and $\dot{\delta}_{c}$
which can be used to to set $\theta_{c}$ in terms of $\phi$ and
$\dot{\delta}_{c}$. We can now substitute this expression for $\theta_{c}$
in Eq.\ (\ref{eq:dot_theta_c}), which becomes a relation between
$\phi$, $\dot{\delta}_{c}$ and $\ddot{\delta}_{c}$, and which can
now be used to set the value of $\phi$. Finally we can substitute
the expressions of $\theta_{c}$ and $\phi$ in Eq.\ (\ref{eq:eom_0}),
as to find a second order closed differential equation for $\delta_{c}$.
On studying the behavior of such equation in the high-$k$ limit,
we find it can be written as 
\begin{equation}
\ddot{\delta}_{c}+aH\dot{\delta}_{c}-\frac{3}{2}\,\frac{G_{{\rm eff}}}{G_{N}}\,\varrho_{c}\,a^{2}\,\delta_{c}=0\,,
\end{equation}
so that the friction term becomes standard, i.e.\ $aH$, while the
mass term acquires a non-standard value, which can be written as 
\begin{equation}
\frac{G_{{\rm eff}}}{G_{N}}=\frac{2}{2-Y\theta}-\frac{3Y\theta\Omega_{m}}{(Y\theta-2)^{2}}+\frac{2\eta_{X}\epsilon_{X}Y}{(Y\theta-2)^{2}}\,.
\end{equation}
This result recovers the standard case in the smooth limit $\theta\to0$
(or at early times, i.e. the limit $Y\to0$). Since we also set constraints
coming from the Integrated Sachs Wolfe effect, we need to find the
relation between $\psi_{{\rm ISW}}\equiv\phi+\psi$ and the matter
density profile $\delta_{c}$, as in 
\begin{equation}
\psi_{{\rm ISW}}=-\frac{3H_{0}^{2}\Omega_{m0}}{k^{2}}\,\frac{\Sigma\,\delta_{c}}{a}\,.
\end{equation}
In the above mentioned procedure to find $G_{{\rm eff}}/G_{N}$, we
have found both $\phi$ and $\psi$ in terms of $\delta_{c}$ and
its time derivatives. Therefore, it is straightforward to find in
the high-$k$ limit the result 
\[
\Sigma=\frac{8+[2\eta_{X}\epsilon_{X}-(4+3\Omega_{m})\theta]\,Y}{2(Y\theta-2)^{2}}\,,
\]
which reduces to unity, as expected, in the $\Lambda$CDM limit.

\subsection{No ghost conditions and stability}

We know the theory does not introduce any new gravity degree of freedom
besides the standard matter fields, whose action have not been modified.
It is then expected that the no-ghost conditions and the no-Laplacian-instability
conditions are trivially satisfied by matter fields. In fact, on reducing
the degrees of freedom as we do in $\Lambda$CDM it is possible to
reach a minimal, reduced Lagrangian only for the propagating degrees
of freedom, say $\delta_{I}$, out of which, in the high-$k$ regime,
we immediately find the following two stability conditions 
\begin{eqnarray}
Q_{I} & = & \frac{\rho_{I}^{2}}{(\rho_{I}+P_{I})}\,\frac{a^{2}}{k^{2}}>0\,,\\
c_{s,I}^{2} & = & \frac{\dot{P}_{I}}{\dot{\rho}_{I}}\geq0\,,
\end{eqnarray}
which are exactly the same conditions also found in $\Lambda$CDM,
and which are trivially satisfied by standard matter fields. In particular
no ghost degrees of freedom are present in the theory (see also \cite{Felice_2016}).

\subsection{Tensor modes}

In order to study the tensor modes for this theory we first set the
perturbation variables as follows 
\begin{equation}
\gamma_{ij}=a^{2}\,(\delta_{ij}+h_{ij})\,,
\end{equation}
where $h_{ij}=\sum_{f={+},{\times}}\epsilon_{ij}^{f}\,h_{f}$, and
$f$ runs over the two polarization modes. The two three-dimensional
symmetric matrices $\epsilon_{ij}^{f}$ satisfy the traceless and
transverse properties, namely $\delta^{ij}\epsilon_{ij}^{f}=0$ and
$\epsilon_{il}^{f}\delta^{lj}\partial_{j}h_{f}=0$, and the normalization
condition holds $\delta^{ij}\epsilon_{ik}^{{+}}\delta^{kl}\epsilon_{lj}^{{+}}=1=\delta^{ij}\epsilon_{ik}^{{\times}}\delta^{kl}\epsilon_{lj}^{{\times}}$,
together with $\delta^{ij}\epsilon_{ik}^{{+}}\delta^{kl}\epsilon_{lj}^{{\times}}=0$.
We are now ready to expand the action at second order in terms of
these tensor modes perturbation variables. We arrive at an action
which leads to a trivial and positive no-ghost condition but differs
from the one of $\Lambda$CDM in the mass term, as expected from a
theory of massive gravity. In fact, the equation motion for each perturbation
can be easily found and can be written as follows 
\begin{equation}
\ddot{h}_{f}=-2\,\frac{\dot{a}}{a}\,\dot{h}_{f}-(k^{2}+\mu^{2}\,a^{2})\,h_{f}\,,
\end{equation}
where $f\in\{{+},{\times}\}$, and 
\begin{equation}
\mu^{2}=\frac{H_{0}^{2}\,[(\theta^{2}Y-2\eta_{X})\,\epsilon_{X}+4\theta]}{4}\,,
\end{equation}
so that when $\epsilon_{X}\to0$, that is when $X={\rm constant}$,
then $\theta=\mu^{2}/H_{0}^{2}={\rm constant}$, as expected. However,
in general $\mu^{2}$ is a function of time.

\section{Confrontation with the data}

Now that we have described all the necessary theoretical changes to
implement in the Boltzmann solver, we want to discuss here the chosen
data sets which will be used in order to constrain the parameters
of the theory. We will make use of both early and late time (high
and low redshifts) cosmological data. 
\begin{itemize}
\item We will make use of the Planck 2018 data \cite{2020}, in particular
we will make use of low-$l$ data ($l\leq29$), and for $l\geq30$,
we will consider temperature (TT), polarization (EE) power spectra,
together with cross correlation of temperature and polarization (TE).
The influence of MTMG on these observables is non trivial, but we
will see that the parameter space will actually be strongly constrained. 
\item BAO (baryon acoustic oscillations) connects fluctuations of baryonic
matter with acoustic waves of the primordial plasma (which can travel
until recombination time). We will make use in particular of the data
compilation in \cite{Alam_2021}. 
\item We make use of the RSD data compilation of ``Gold-2018'' data set
considered in \cite{Sagredo_RSD}, and the likelihood developed in
\cite{Arjon_2020}. This data set is particularly sensitive to the
growth of perturbation which is sourced in turn in this theory by
a modified value for the quantity $G_{{\rm eff}}/G_{N}$. In terms
of the equations of motion we have, this will affect the power spectrum
for the matter perturbations. 
\item We make use also of the Pantheon data (without implementing any prior
on the absolute magnitude for the Type Ia supernovae), accordingly
to \cite{Scolnic:2017caz}. This data set constrains the dynamics
of the universe at large scale in the following redshift window, $0.01\leq z\leq2.3$. 
\item ISW-galaxy cross correlation data have been used extensively in order
to constrain dark energy models. Although this theory does not add
any extra degree of freedom in the gravity (or matter) sector, still
the function $\Sigma$ gets modified from unity and this in turn will
set strong constraints for MTMG. In particular we will use the catalogs
(2MASS Photometric Redshift catalog, the WISE\texttimes SuperCOSMOS
photo-z catalog, the NRAO VLA SkySurvey radio sources catalog, the
SDSS DR12 and SDSSDR6 QSO photometric catalogs) and a modified version
of the likelihood (according to the new dynamics for $\Sigma$) presented
in \cite{St_lzner_2018}. 
\end{itemize}
We have not inserted here several possibly interesting other data
sets for several reasons. First of all, since there is a large discrepancy
among several experiments for the same observable, $H_{0}$, we have
chosen not to use data priors giving some constraint on this variable.
To choose any prior for it would imply being sure about the systematics
of all the experiments which are now running to understand its value,
a knowledge which is at the moment missing at least for us. Nonetheless
the remaining data sets will already constrain its value which will
correspond to a prediction of the theory for it. On top of that we
will present predictions for the theory also for the value of $S_{8}=\sqrt{\Omega_{m0}/0.3}$,
and we will check its value compared to the latest results of both
KiDS and DES.

We will make use of both modified code for CLASS and Montepython \cite{Blas:2011rf,Audren:2012wb,Brinckmann:2018cvx},
as to adjust the theory of MTMG, in order to find, via a Metropolis-Hastings
sampling algorithm, the available parameter space compatible with
the given data sets.

\section{Results}

We will consider here the study of MTMG from two different approaches
which in our intentions will complement each other. We find it useful
to separate the contribution coming from the Planck data as to see
whether in the context of MTMG there is a tension in the data. This
procedure, on the other hand, will make it clear the contribution
from early-time high-precision data to the constraint on the mass
of the graviton which is expected in this theory to be of order of
$H_{0}$.

\subsection{Data without including Planck}

First of all, in the normal branch of MTMG, we have as one of the
predominant and peculiar effects the fact that $G_{{\rm eff}}/G_{N}$
can deviate from unity. This immediately opens the possibility of
having weak gravity implemented in the theory. In order to address
this issue we then study the RSD data only. The results of this study
can be read in Table \ref{tab:main_results_RSD}. In particular, this
measurement is not able by itself to constrain the theory much and
allows for a large parameter space, especially (but not only) for
negative values of $\theta_{0}$.

\begin{table}
\begin{tabular}{ccccc}
\hline 
data set  & $\ensuremath{\theta_{0}}$  & $\ensuremath{\Omega_{m}}$  & $\ensuremath{\sigma_{8}}$  & $\ensuremath{S_{8}}$\tabularnewline
\hline 
RSD  & $\ensuremath{-1800_{-9900}^{+1852}}$  & $\ensuremath{0.65_{-0.50}^{+0.55}}$  & $\ensuremath{1.28_{-0.59}^{+1.1}}$  & $\ensuremath{2.0_{-1.4}^{+2.3}}$\tabularnewline
RSD+BAO+Pantheon  & $\ensuremath{-1.6_{-4.5}^{+3.1}}$  & $\ensuremath{0.296_{-0.027}^{+0.030}}$  & $\ensuremath{0.804_{-0.086}^{+0.098}}$  & $\ensuremath{0.798_{-0.090}^{+0.11}}$\tabularnewline
RSD+BAO+Pantheon+ISW  & $\ensuremath{-0.55_{-1.1}^{+0.91}}$  & $\ensuremath{0.293_{-0.018}^{+0.018}}$  & $\ensuremath{0.777_{-0.051}^{+0.052}}$  & $\ensuremath{0.768_{-0.056}^{+0.055}}$\tabularnewline
\hline 
\end{tabular}

\caption{Constraints at 95\% CL on $\theta_{0}=\mu_{0}^{2}/H_{0}^{2}$, $\Omega_{m}$,
$\sigma_{8}$, and $S_{8}$ as inferred from different combinations
of data sets (but without Planck), in the MTMG model.\label{tab:main_results_RSD}}
\end{table}

\begin{figure}
\includegraphics[width=9cm]{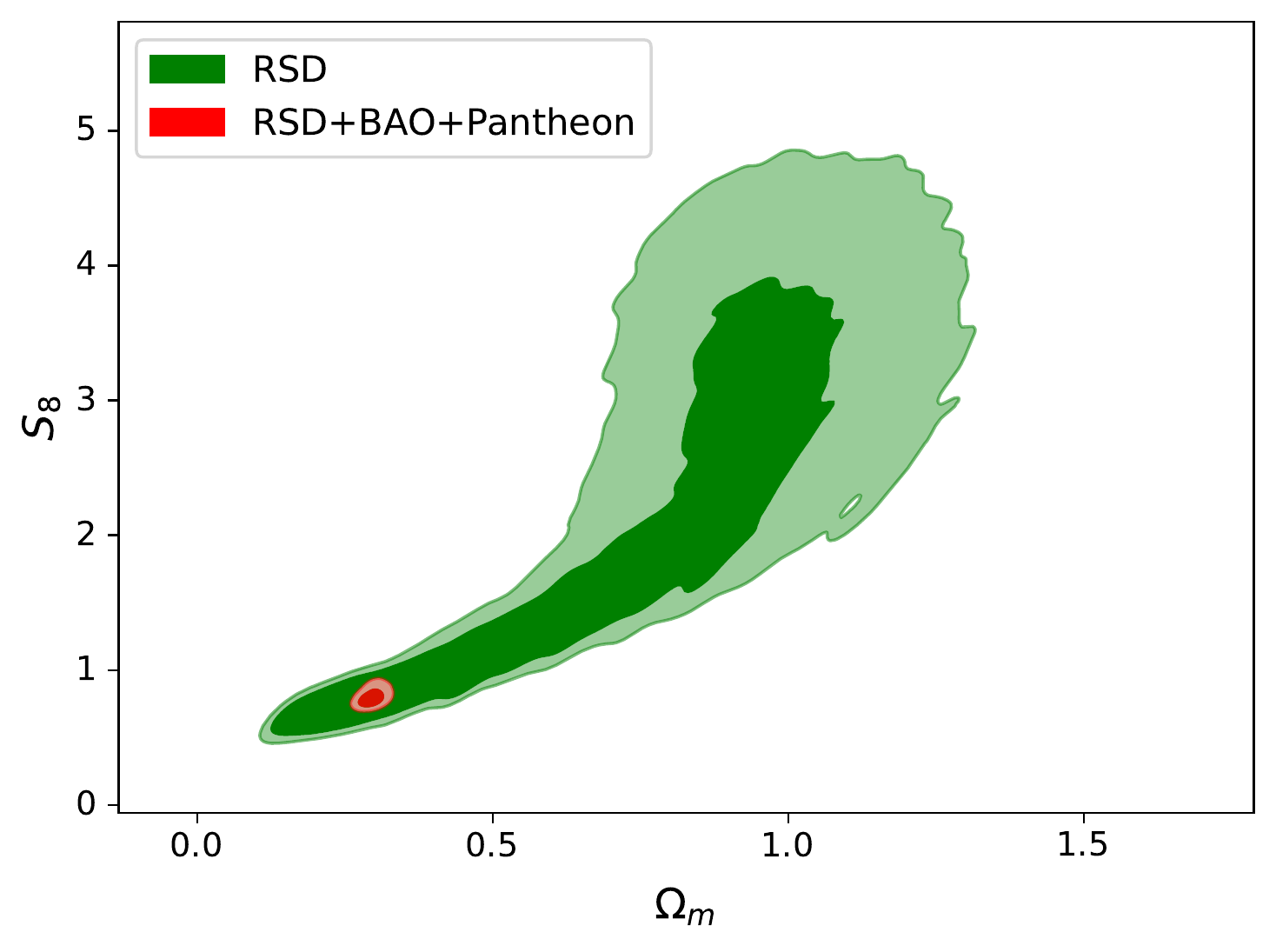}

\caption{We show here the large degeneracy in parameter space the RSD data
allow for the MTMG model and how instead BAO and Pantheon data effectively
reduce it.\label{fig:RSD-=00003D000026-RSD+BAO+P}}
\end{figure}

We have studied first the influence of the RSD data alone on MTMG.
We find that there is a large degeneracy in the parameter space allowed
by this single data set, see e.g.\ in Fig.\ \ref{fig:RSD-=00003D000026-RSD+BAO+P}
the large parameter space for $S_{8}$ and how it reduces when other
data sets are included. In particular, from Table \ref{tab:main_results_RSD},
we can see that the RSD data alone allow for both positive and even
quite large negative values for $\theta_{0}$, meaning that this data
set is compatible with a tachyonic graviton having a negative squared
mass. To have a tachyonic graviton implies the tensor modes having
energy of order $E_{{\rm GW}}\simeq\sqrt{|\theta_{0}|}H_{0}$, that
is with a wave-length much larger than the typical gravitational waves
produced astrophysically, will start showing an instability in a cosmological
time scale $\propto E_{{\rm GW}}^{-1}$.

\begin{figure}
\includegraphics{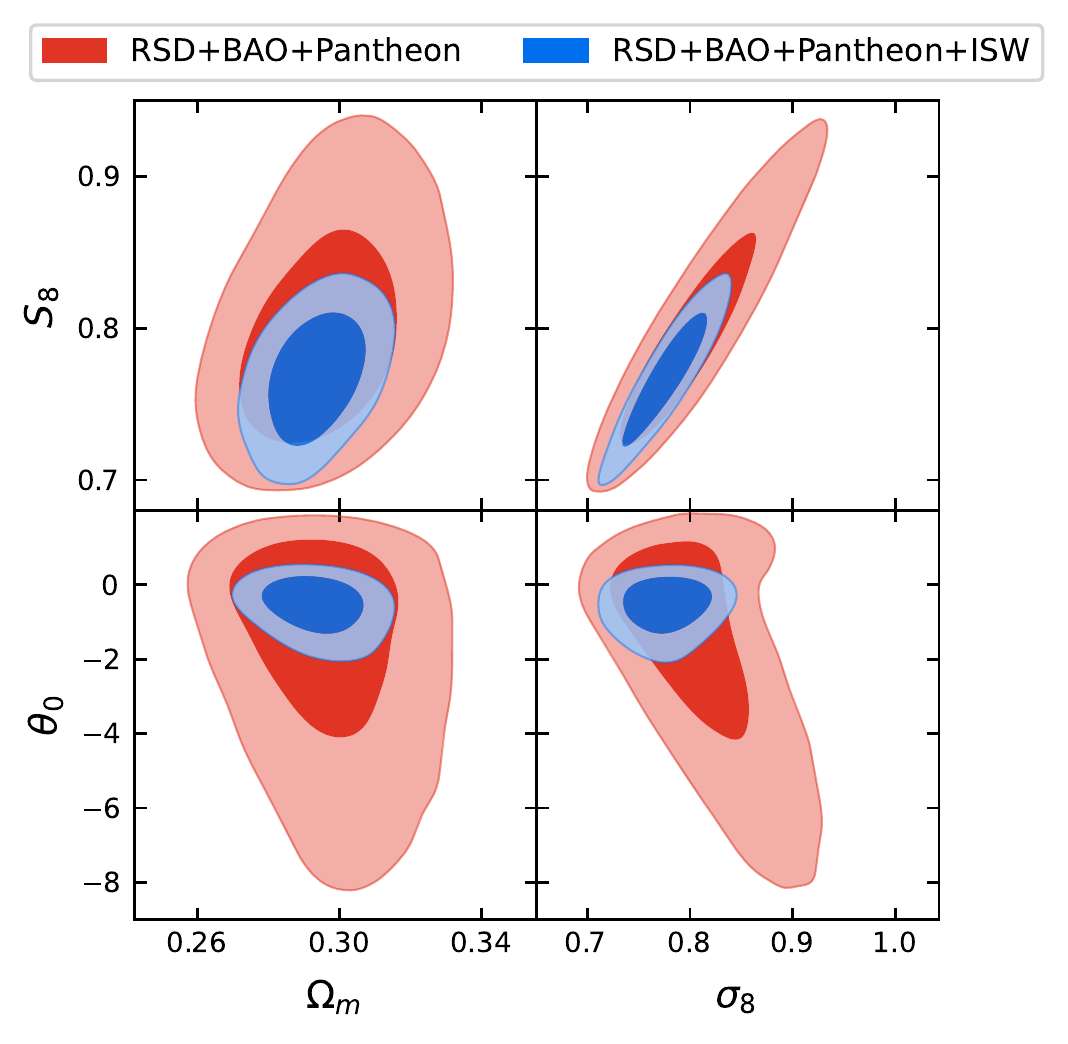}

\caption{We show here that MTMG is sensitive also to ISW data, as the allowed
parameter space further shrinks after we make use of them.\label{fig:RSD+BAO+P-VS-LateTimes}}
\end{figure}

On the other hand, if we also add the both BAO and Pantheon data the
allowed parameter space considerably shrinks. Finally, on adding the
ISW data, degeneracy further breaks down into a smaller region presented
in Fig.\ \ref{fig:RSD+BAO+P-VS-LateTimes}. 

It should be noticed that at 68\% CL, we have $\theta_{0}=\ensuremath{-0.55_{-0.37}^{+0.61}}$,
which clearly states that: 1) $\Lambda$CDM is inside the 1-$\sigma$
level and 2) positive values for the mass of the graviton are well
inside this allowed range of values. Now we can move on to study whether
and how Planck will influence the final results.

\subsection{Data including Planck}

We are now ready to analyze MTMG in the light of Planck data. Let
us remind the reader that we have to sample a $\chi^{2}$ which is
a function of eleven free parameters, namely 
\begin{equation}
\chi^{2}=\chi^{2}(\omega_{b},\omega_{{\rm cdm}},\theta_{s},A_{s},n_{s},\tau_{{\rm reio}},A_{1},A_{2},\bar{c}_{1},\bar{c}_{2},\bar{c}_{3})\,,
\end{equation}
where the first six ones are the same as in $\Lambda$CDM. The remaining
five parameters are the new parameters introduced in MTMG in order
to give a quite large class of possibilities in dynamics\footnote{We have checked that $\chi^{2}$ does not improve on adding $A_{3}$
as a free parameter. This leads to the conclusion that the data are
not sensitive enough to the way how the variable $X=\tilde{a}/a$
changes with time between the two asymptotic constant values. This
leads us to believe that the constraints on the graviton mass found
here will hold for other different dynamical choices for $X(t)$.}. Having eleven parameters, five more than $\Lambda$CDM, would give
the idea that the constraints we obtains will not be so tight. This
expectation is only partially true. Actually, the mass of the graviton,
which is a function of the MTMG parameters and which is a derived
parameter, directly and strongly affects the behavior of gravity and
the structure formation (i.e.\ perturbations) so that, in fact, we
obtain what we think is the strongest constraint on the mass of the
graviton for the normal branch of MTMG.

\begin{figure}
\subfloat[Constraints/predictions on the variable $S_{8}$.]{\includegraphics[width=8cm]{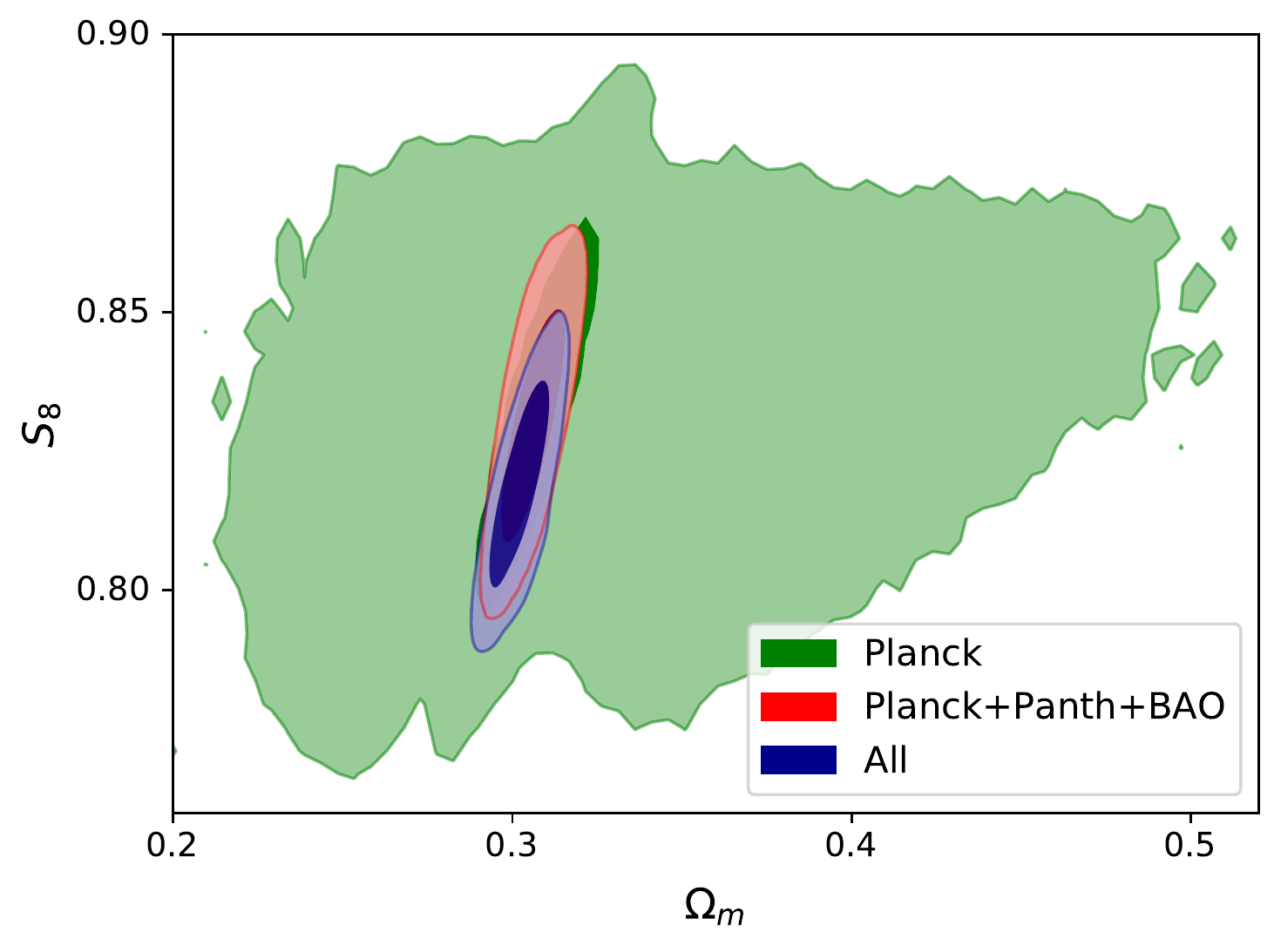}

}\hfill{}\subfloat[Constraints on $\theta_{0}=\mu_{0}^{2}/H_{0}^{2}$, where $\mu_{0}^{2}$
is today's squared mass of the graviton.\label{fig:theta0_planck_BAO}]{\includegraphics[width=8cm]{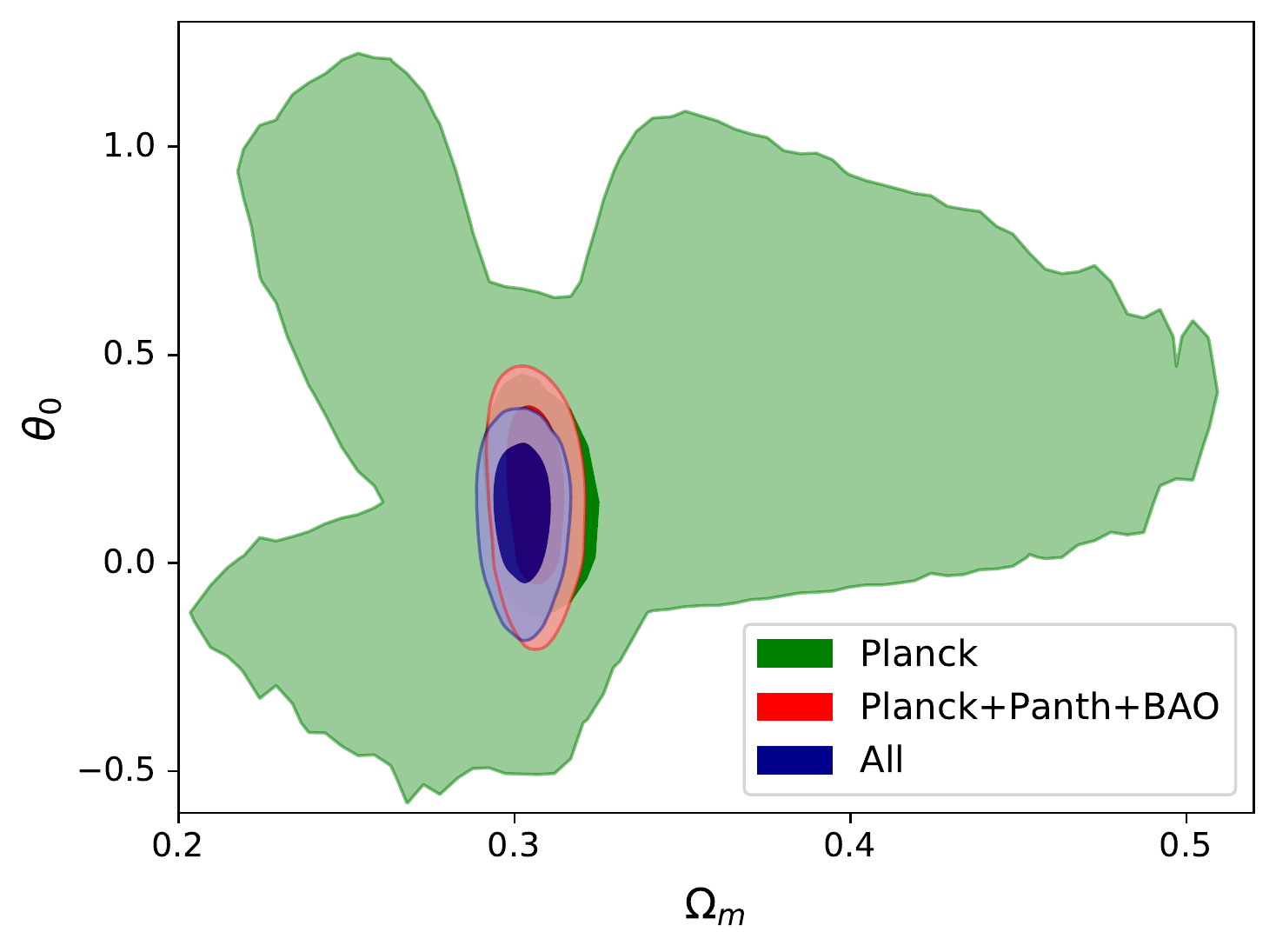}

}\caption{Analysis of Planck data, alone and together with BAO+Pantheon and
further adding RSD+ISW.\label{fig:Analysis-of-Planck}}
\end{figure}

In order to achieve this goal, and to see whether Planck data in the
context of MTMG was suffering from tensions with the other data sets
we have considered here, we have first of all made a run only using
Planck data. The results for all the parameters of interest are shown
in Table \ref{tab:Constraints-at-95-all} and Fig.\ \ref{fig:Analysis-of-Planck}.
In particular, in Fig.\ \ref{fig:theta0_planck_BAO}, we can see
that, thanks to Planck data, both positive and negative values for
$\theta_{0}$ are still allowed but a large chunk of the negative
parameter space is now not available any longer.

\begin{figure}
\includegraphics[width=9cm]{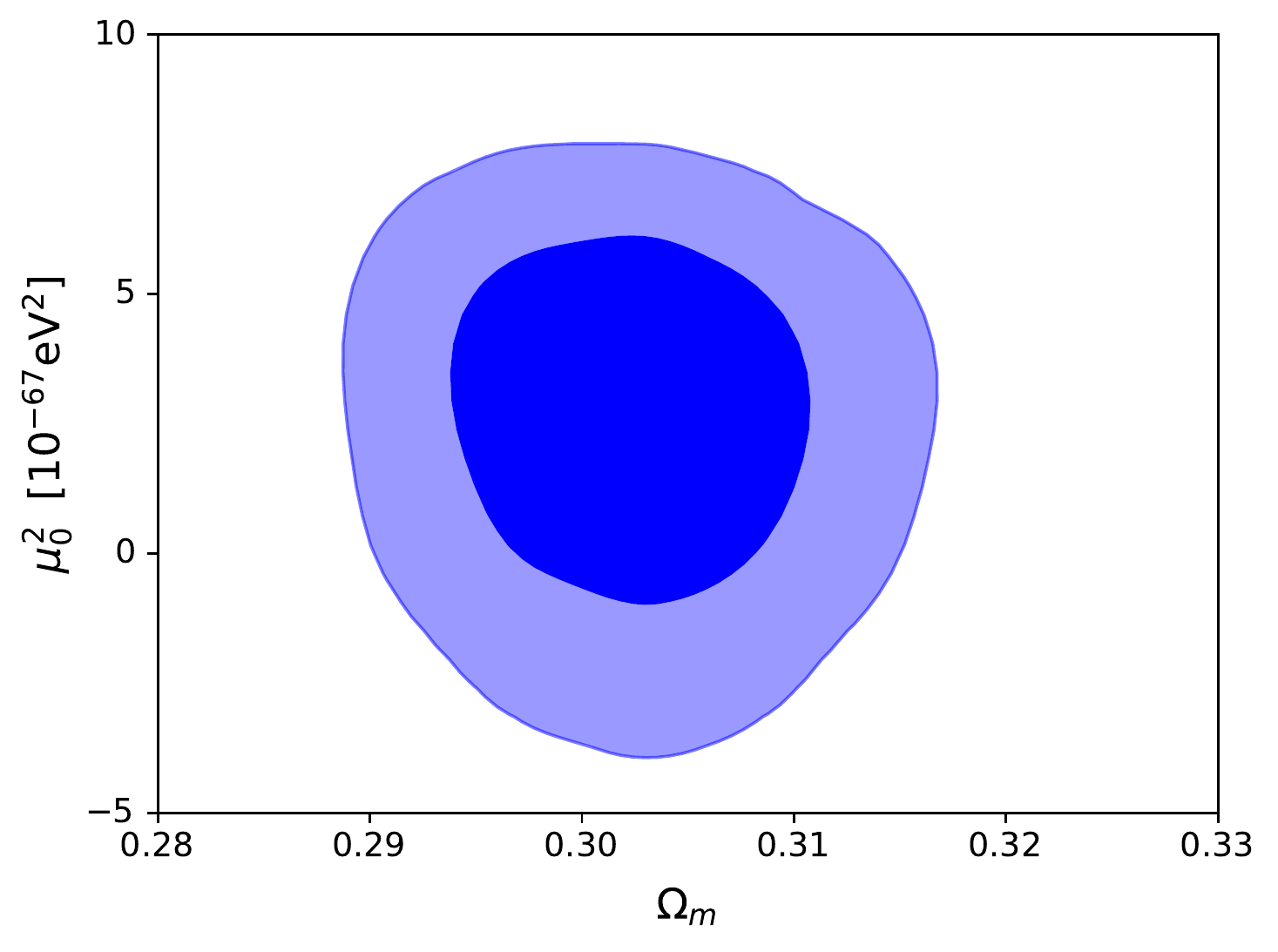}

\caption{Constraints on $\mu_{0}^{2}$ for the joint analysis.\label{fig:Constraints-on-mu2}}
\end{figure}

\begin{figure}
\subfloat[Constraints on the variable $S_{8}$ as a function of $H_{0}$ in
MTMG.]{\includegraphics[width=8cm]{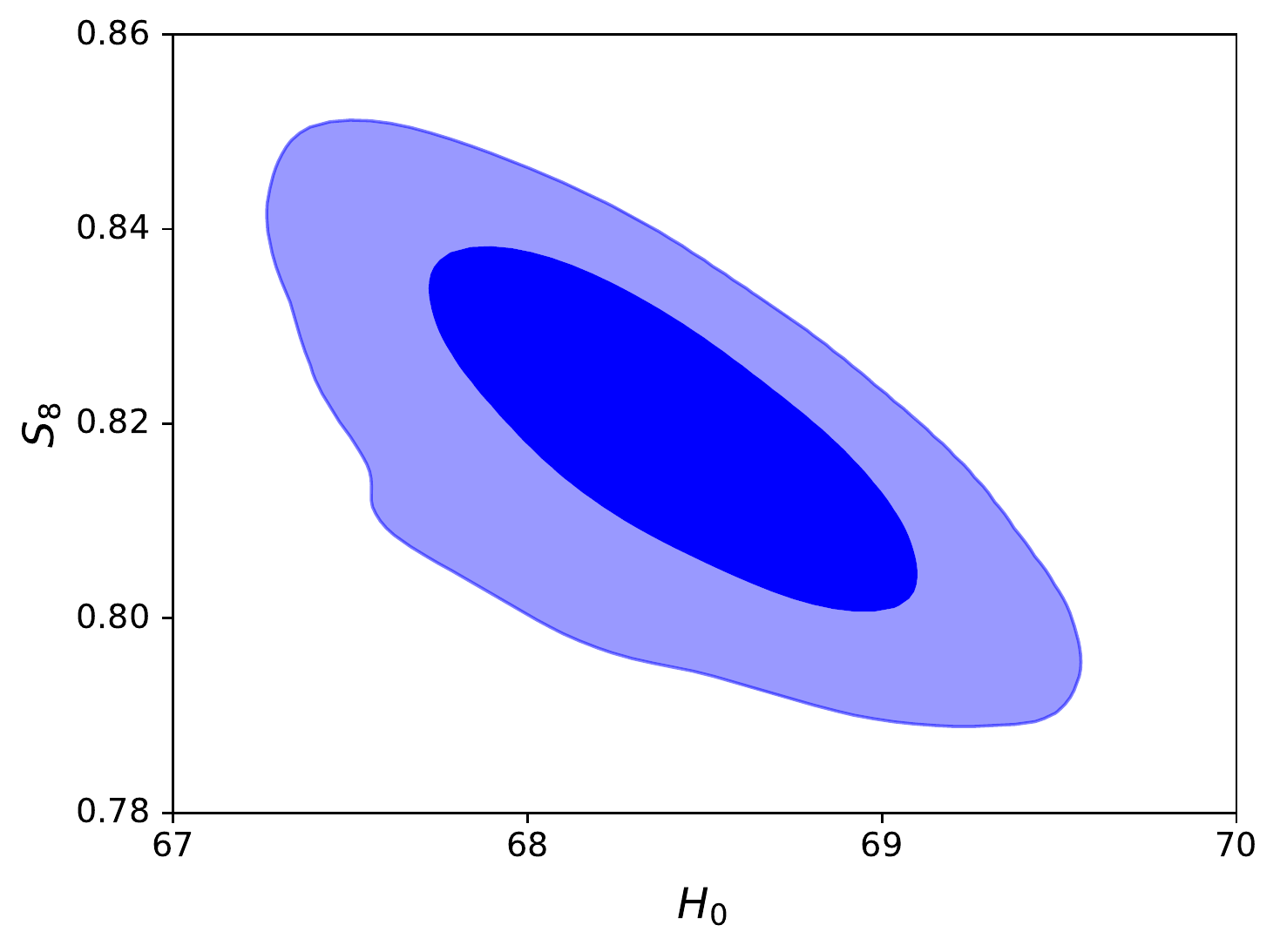}

}\hfill{}\subfloat[Constraints on $\theta_{0}=\mu_{0}^{2}/H_{0}^{2}$ as a function of
$H_{0}$ in MTMG.]{\includegraphics[width=8cm]{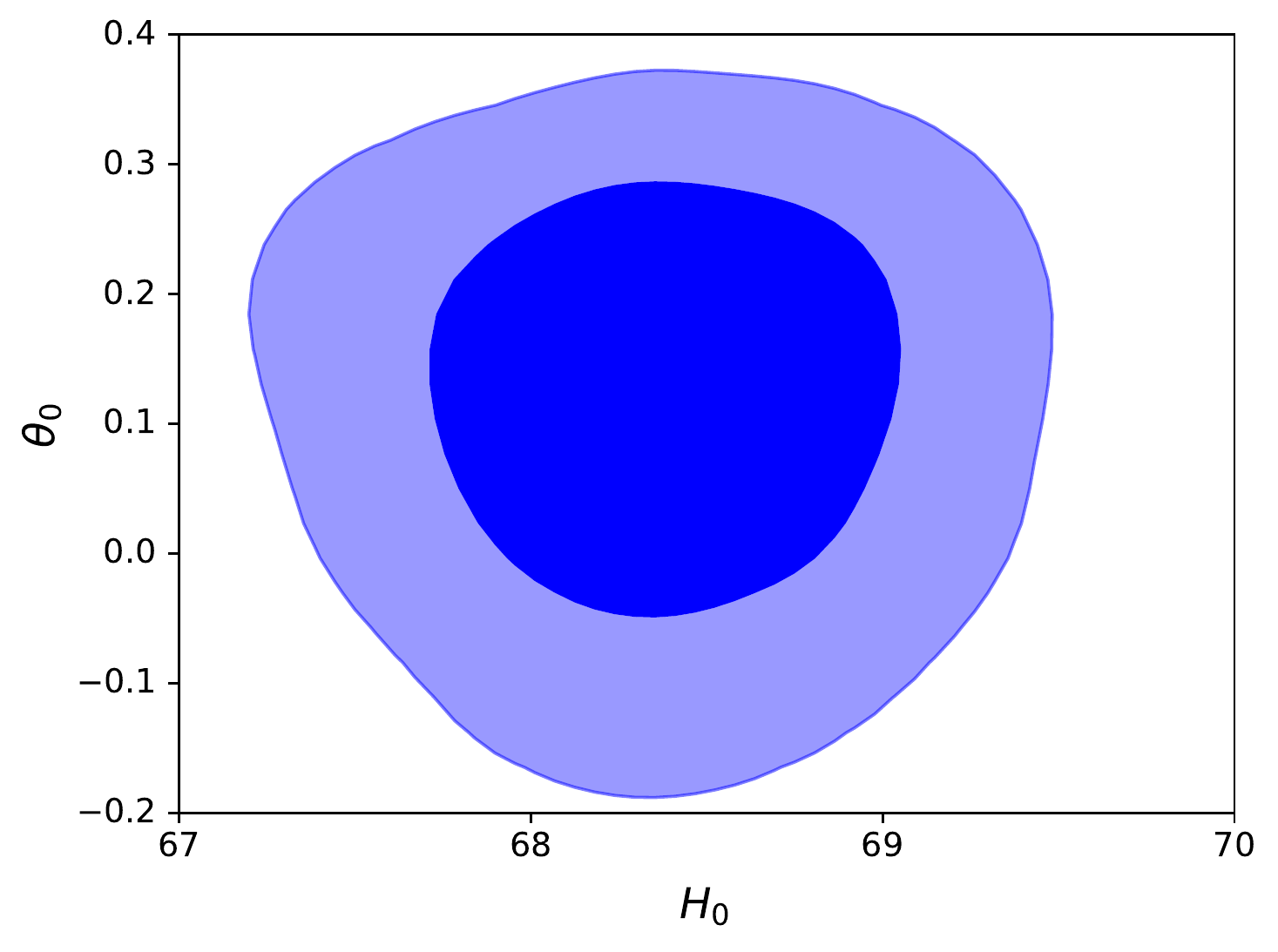}

}\caption{Predictions given by the theory of MTMG regarding the observables
$\theta_{0}$, $S_{8}$ and $H_{0}$.\label{fig:theta0_S8_H0}}
\end{figure}

In the second and third steps we perform the $\chi^{2}$ adding BAO+Pantheon
and then RSD+ISW as well. After marginalization over the parameters,
for the constraint on $\theta_{0}$ at 68\% CL we find $\theta_{0}=0.119_{-0.098}^{+0.12}$.
This result shows that Planck tends to prefer non-negative values
for the squared mass of the graviton, i.e.\ Planck prefers non-tachyonic
gravitons in MTMG. Since, at 95\% CL, we have $\theta_{0}=0.12_{-0.22}^{+0.21}$,
then we can \emph{exclude} all values for which $\theta_{0}<-0.1$
and $\theta_{0}>0.33$, which does not exclude $\Lambda$CDM. Furthermore,
we find at 95\% CL the following constraint on today's squared mass
for the graviton $\mu_{0}^{2}=2.5{}_{-4.8}^{+4.5}\times10^{-67}\ {\rm eV}^{2}$,
which implies, in particular, an upper bound $\mu_{0}<8.4\times10^{-34}$
eV. We finally show for the joint analysis of all the data the 95\%
CL for $\mu_{0}^{2}$ in Fig.\ \ref{fig:Constraints-on-mu2}. We
add the information on how MTMG is sensitive to the value of today's
Hubble parameter, and how both $S_{8}$ and $\theta_{0}$ are related
to it in Fig.\ \ref{fig:theta0_S8_H0}.

If the constraints on today's value for the mass of the graviton turn
out to be quite stringent (because most of the late time data sets
are sensitive to it, i.e.\ they put constraints on the positive parameter$A_{1}$
on which $\mu_{0}$ also depends), we should expect that the value
of the graviton mass at high redshifts should be less constrained
as only Planck can sets some constraints on it. In fact at 95\% CL
we find $\mu_{{\rm \infty}}^{2}=0_{-69}^{+44}\times10^{-67}{\rm eV}^{2}$.

\begin{table}
\begin{tabular}{cccc}
\hline 
 & Planck  & Planck+BAO+Pantheon  & All joint analysis\tabularnewline
\hline 
$\ensuremath{10^{2}\omega_{b}}$  & $2.242_{-0.030}^{+0.031}$  & $2.242_{-0.027}^{+0.027}$  & $2.247_{-0.027}^{+0.027}$\tabularnewline
$\omega_{{\rm cdm}}$  & $0.1197_{-0.0028}^{+0.0028}$  & $0.1195_{-0.0020}^{+0.0020}$  & $0.1189_{-0.0019}^{+0.0019}$\tabularnewline
$100\theta_{s}$  & $1.04194_{-0.00058}^{+0.00059}$  & $1.04194_{-0.00056}^{+0.00056}$  & $1.04198_{-0.00056}^{+0.00057}$\tabularnewline
$\ln10^{10}A_{s}$  & $3.044_{-0.032}^{+0.032}$  & $3.045_{-0.032}^{+0.033}$  & $3.037_{-0.031}^{+0.031}$\tabularnewline
$n_{s}$  & $0.9671_{-0.0088}^{+0.0090}$  & $0.9674_{-0.0076}^{+0.0077}$  & $0.9683_{-0.0075}^{+0.0074}$\tabularnewline
$\tau_{{\rm reio}}$  & $0.055_{-0.015}^{+0.016}$  & $0.055_{-0.015}^{+0.016}$  & $0.052_{-0.015}^{+0.015}$\tabularnewline
$A_{1}$  & $0.57_{-0.57}^{+1.1}$  & $0.63_{-0.63}^{+0.73}$  & $0.71_{-0.71}^{+0.43}$\tabularnewline
$A_{2}$  & $6.2_{-7.0}^{+8.4}$  & $6.4_{-6.4}^{+8.5}$  & $3.9_{-3.9}^{+11}$\tabularnewline
$\bar{c}_{1}$  & $0.0_{-9.2}^{+9.2}$  & $0.2_{-9.2}^{+9.0}$  & $-0.1_{-8.5}^{+8.3}$\tabularnewline
$\bar{c}_{2}$  & $0.1_{-8.4}^{+8.4}$  & $0.0_{-8.3}^{+8.5}$  & $-0.4_{-7.1}^{+6.8}$\tabularnewline
$\bar{c}_{3}$  & $1.2_{-8.1}^{+8.1}$  & $1.1_{-8.0}^{+8.2}$  & $0.9_{-6.5}^{+6.6}$\tabularnewline
\hline 
$\Omega_{m}$  & $0.318_{-0.068}^{+0.17}$  & $0.306_{-0.012}^{+0.012}$  & $0.302_{-0.011}^{+0.011}$\tabularnewline
$H_{0}$  & $67_{-10}^{+8}$  & $68.11_{-0.92}^{+0.92}$  & $68.37_{-0.93}^{+0.87}$\tabularnewline
$\sigma_{8}$  & $0.816_{-0.15}^{+0.089}$  & $0.822_{-0.018}^{+0.021}$  & $0.816_{-0.017}^{+0.016}$\tabularnewline
$S_{8}$  & $0.832_{-0.040}^{+0.040}$  & $0.830_{-0.027}^{+0.028}$  & $0.819_{-0.024}^{+0.023}$\tabularnewline
$\Delta$  & $-0.4_{-4.2}^{+2.7}$  & $-0.4_{-4.1}^{+2.5}$  & $-0.1_{-1.5}^{+1.3}$\tabularnewline
$\theta_{0}$  & $0.18_{-0.40}^{+0.64}$  & $0.16_{-0.28}^{+0.27}$  & $0.12_{-0.22}^{+0.21}$\tabularnewline
$\bar{c}_{4}$  & $3_{-10}^{+10}$  & $3_{-11}^{+11}$  & $3.2_{-6.9}^{+5.9}$\tabularnewline
\hline 
\end{tabular}

\caption{Constraints at 95\% CL on the primary and derived parameters of dynamical
MTMG.\label{tab:Constraints-at-95-all}}
\end{table}


\section{Conclusions}

The minimal theory of massive gravity (MTMG) is a theory with only
two gravitational degrees of freedom which reproduces the dRGT massive
gravity background for a homogeneous and isotropic universe without
leading to ghost or strong coupling in general. As reminiscent of dRGT, the theory 
on a cosmological background allows for the presence of two branches,
which are defined by the following background constraint (which comes
in addition to the modified Einstein equations): 
\begin{equation}
(c_{1}X^{2}+2c_{2}X+c_{3})\left(\frac{\dot{X}}{N}+HX-H\,\frac{M}{N}\right)=0\,,\label{eq:branches}
\end{equation}
where $N$ and $M$ are the physical and fiducial lapse respectively,
$c_{i}$ ($i=1,2,3$) are free parameters of the theory, $H=\dot{a}/(aN)$
is the Hubble expansion rate, whereas $X=\tilde{a}/a$ is the ratio
between the fiducial and the physical scale factors.

The self-accelerating solution, the one which fixes $X$ to be the
constant satisfying $c_{1}X^{2}+2c_{2}X+c_{3}=0$, is indistinguishable
from $\Lambda$CDM except for the tensor modes dynamics which acquire
a non-zero mass. Therefore for this self-accelerating branch we can
only assume the standard graviton-mass bounds to hold and they are
trivially fulfilled once we impose from cosmological purposes that
such a graviton mass is of order of $H_{0}$. In this latter case the self-accelerating branch of 
MTMG is trivially consistent with all experiments and observations, 
leaving no detectable imprint in the present data.

On the other hand, for the normal branch of MTMG, we need the equation
$\frac{\dot{X}}{N}+HX-H\,\frac{M}{N}=0$ to hold. Since the fiducial
metric (in the unitary gauge) corresponds to a given external field
(that we fix to be homogeneous and isotropic, and only time-dependent,
as to respect the symmetry of the FLRW background), it does not have
its own dynamics and is a part of the definition of the
theory. In particular on giving $X=X(z)$, Eq.\ (\ref{eq:branches})
then uniquely fixes $M$ and from this point onward the theory will
have its own phenomenological predictions which can be tested against
the data.

We have thus studied the normal branch of MTMG in order to understand
its predictions against some of the most recent and updated cosmological
data. Although this in principle means to study the functional freedom
of $X(t)$, in practice we have found it convenient to consider a
simple model which describes a smooth three-parameters transition
between two different values of $X$. In other words this model interpolates
between two different values of the mass of the graviton, one describing
the mass at high redshifts and the other one corresponding to today's
value for the mass of the graviton. Although this transition is still
a special dynamics among all the possible ones, we think it captures
the observationally relevant possibilities of the whole MTMG model
for several reasons: (i) the background corresponds to a transition
between two $\Lambda$CDM backgrounds (which differ by their effective
cosmological constant), and we know that $\Lambda$CDM has still an
extremely good fit to the data\footnote{We have not considered here the $H_{0}$-tension as several measurements
still give different answers in different methodologies, leading to
as yet an obscure understanding of the phenomenon. As for the $S_{8}$
tension instead, most of the constraints we have are found on assuming
$\Lambda$CDM to hold at all times, so it becomes rather difficult
to compare the predictions of MTMG with the (interesting) results
from both KiDS and DES surveys.}, and, although not impossible, it is quite hard to outperform it
(see e.g.~\cite{DeFelice:2020cpt}); (ii) the final results on today's
value of the graviton mass $\mu_{0}$ closely agree with the ones
recently shown in \cite{deAraujo:2021cnd} for the simplest possible
choice that $X(t)={\rm constant}$, namely the case where the background (but not
the perturbations) is exactly the same as in $\Lambda$CDM. For these
reasons we believe that the constraints we have found for the today's
graviton mass for the normal branch of MTMG will hold even for other
dynamics of $X(t)$, after we impose all the constraints coming from
the data.

After implementing the full Boltzmann equations for the perturbations,
and on studying several data sets (and several combinations of them)
which include Planck 2018, Pantheon (Supernovae Type Ia), baryonic
acoustic oscillations (BAO), redshift space distortion (RSD), and
ISW-galaxy correlation data sets, we have arrived at the following
conclusions. First of all, inside these data sets MTMG does not feel
any internal tension, so that all the data sets constrain the model
in the most efficient way. Furthermore, although we use Planck 2018
and even if we have five new parameters (compared to $\Lambda$CDM),
still we find strong constraint on today's value of the graviton mass,
to which most of the late-time data are sensitive. We then find that,
at 95\% CL, today's squared mass for the graviton is bound to be in
the following range: $\mu_{0}^{2}=2.5{}_{-4.8}^{+4.5}\times10^{-67}\ {\rm eV}^{2}$,
see also Fig.\ \ref{fig:Constraints-on-mu2}.

This result does not evidently rule out $\Lambda$CDM (for which $\mu_{0}$
vanishes), but rather, it gives a very interesting upper bound for
the graviton mass in the normal branch of MTMG as $\mu_{0}<8.4\times10^{-34}$
eV, which is the strongest so far for an existing theory of massive
gravity. We hope this result will motivate further new studies in
the field of massive gravity from both theoretical and phenomenological
sides.
\begin{acknowledgments}
  The work of A.D.F.\ was supported by Japan Society for the Promotion
  of Science Grants-in-Aid for Scientific Research No.\
  20K03969. S.~M.'s work was supported in part by Japan Society for
  the Promotion of Science Grants-in-Aid for Scientific Research No.\
  17H02890, No.\ 17H06359, and by World Premier International Research
  Center Initiative, MEXT, Japan. The work of M.C.P.\ was supported by
  the Japan Society for the Promotion of Science Grant-in-Aid for
  Scientific Research No. 17H06359, and also acknowledges the support
  from the Japanese Government (MEXT) scholarship for Research Student
  during the initial phase of this project. The numerical computation
  in this work was carried out at the Yukawa Institute Computer
  Facility.
\end{acknowledgments}

\bibliographystyle{apsrev4-2}
\bibliography{mtmg}

\end{document}